\documentclass[aps,pre,10pt,twocolumn]{revtex4-1}
\usepackage{graphicx}
\usepackage{epstopdf}
\usepackage{xcolor}
\usepackage{amsmath}
\usepackage{amssymb}
\usepackage[utf8]{inputenc}
\usepackage{graphicx,float,hyperref}
\usepackage{natbib}
\usepackage{graphicx,latexsym} 

\DeclareMathOperator{\csch}{csch}

\begin{document}


\title{Relativistic Quantum Otto Engine: Generalized efficiency bounds and frictional effects}

\author{Vahid Shaghaghi}
\affiliation{ Department of Optics, Palacky University, 17. listopadu 1192/12, 779 00 Olomouc, Czech Republic}

\author{Pritam Chattopadhyay}
\affiliation{Department of Chemical and Biological Physics,
Weizmann Institute of Science, Rehovot 7610001, Israel}

\author{Vijit V. Nautiyal}
\affiliation{School of Education, University of New England, Armidale, New South Wales 2350, Australia }

\author{Kaustav Chatterjee}
\affiliation{
Center for Macroscopic Quantum States (bigQ), Department of Physics, Technical University of Denmark, \\
 Building 307, Fysikvej, 2800 Kongens Lyngby, Denmark
}%
\author{Tanmoy Pandit}
 \affiliation{Institute for Theoretical Physics, Leibniz Institute of Hannover, Hannover, Germany}
\affiliation{Institute of Physics and Astronomy, TU Berlin, Berlin, Germany}
\author{Varinder Singh}
\email{varinderkias@kias.re.kr}
\affiliation{School of Physics, Korea Institute for Advanced Study, Seoul 02455, Korea}
%
 
%


\begin{abstract}


This work investigates a relativistic quantum Otto engine with a harmonic oscillator as its working medium, analyzing how relativistic motion and nonadiabatic driving affect its performance and efficiency bounds. In the adiabatic regime, a closed-form analytical expression is derived for the generalized Carnot efficiency, which incorporates the effects of relativistic motion and reduces to the standard Carnot efficiency in the nonrelativistic limit. For nonadiabatic driving, we consider sudden compression and expansion work strokes and show that the maximum efficiency achievable by the engine is  limited to 1/2, even in the ultra-relativistic limit. Going one step further, we also derive an analytical expression for the efficiency bound in the sudden-switch protocol, which can be regarded as the nonadiabatic counterpart of the generalized Carnot efficiency. Together, these results provide analytical bounds for the efficiency of relativistic quantum heat engines and constitute the first systematic study of the interplay between relativistic motion and frictional effects arising from nonadiabatic driving.
  

\end{abstract}


\maketitle 

%
\section{Introduction}

The study of thermal machines lies at the heart of both classical and quantum thermodynamics~\cite{SV2016,millen2016perspective,Scully2001,scully_pnas,PhysRevE.87.042131,PhysRevLett.98.240601,PhysRevLett.93.140403,chattopadhyay2021bound,PhysRevLett.105.150603,deffner2018efficiency,chattopadhyay2020non,reviewPr,campisi2016power,santos2023pt,VSatnam2022,singh2020optimal,Kaur2025,VS2020,bera2024steady,mohan2025coherent,VS2023}. From the steam engines that powered the industrial revolution to nanoscale devices at the frontier of quantum technologies, the central objective has remained the same: the conversion of heat into useful work. In classical thermodynamics, performance is constrained by the Carnot bound, $\eta_C=1-T_c/T_h$, which sets the maximum achievable efficiency between hot and cold reservoirs at temperatures 
 $T_h$ and $T_c$. With the rapid development of quantum thermodynamics~\cite{SV2016,Mahler,DeffnerBook,AlickiKosloff}, numerous studies have shown that the classical Carnot bound can be surpassed by quantum heat engines operating with nonclassical resources. Examples include quantum coherence~\cite{Scully2001,ScullyAgarwal2003,Skrzypczyk2014}, quantum correlations~\cite{MNBera,Park2013,Marti2015,OzgurCorrelation}, and squeezed thermal reservoirs~\cite{Lutz2014,Rezek2017,Bijay2017,Klaers2017,Alonso2014,Wang2019,VOzgur2020}, among others~\cite{Ghosh2017}. In these cases, the second law of thermodynamics must be reformulated to incorporate quantum effects, giving rise to generalized Carnot bounds whose precise form depends on the specific quantum resource being exploited ~\cite{MNBera,Niedenzu2018,Obinna2014,Lutz2014}.



However, most studies of quantum heat engines have traditionally been carried out in nonrelativistic regimes. Recently, there have been growing interest in the relativistic quantum thermodynamics \cite{Bruschi2020,FerreriBruschi2023,PapadatosA,Fuentes2014}.
 Here, the Unruh effect provides a natural bridge between relativity, quantum field theory, and thermodynamics, offering a mechanism by which uniformly accelerated detectors perceive the Minkowski vacuum as a thermal reservoir \cite{Unruh1976,Dewitt1979}. Building on this idea, Arias et al.~\cite{Arias2018} proposed the Unruh quantum engine, where the working substance is a qubit modeled as an accelerated Unruh–DeWitt detector that undergoes an Otto cycle, thereby extracting work from the quantum vacuum, which in its accelerated frame manifests as a thermal bath through the Unruh effect.
This construction is widely regarded as the prototype of relativistic quantum heat engines. Subsequent works have extended the Unruh quantum Otto engine to a range of scenarios, including fermionic fields \cite{Gray2018}, degenerate detectors \cite{Xu2020}, entangled detectors \cite{KaneMajhi2021,BarmanMajhi2022}, alternative working media such as qutrit detectors \cite{Hirotani2025}, instantaneous detector–field interactions \cite{KollasMoustos2024,GallockYoshimura2024},  setups with reflecting boundaries \cite{Mukherjee2022}, and thermal machines operating in curved spacetime backgrounds \cite{MoustosAbah2025,misra2024black,ferketic2023boosting}. More broadly, the framework has also been applied to Unruh–DeWitt detectors following arbitrary timelike trajectories in curved spacetime \cite{GallockYoshimura2023}.


Thus far, research on relativistic quantum heat engines has mainly focused on demonstrating the feasibility of work extraction and devising operating protocols. Only very recently have a few studies shown that relativistic motion itself can serve as a genuine quantum resource, enabling efficiencies beyond the classical Carnot bound \cite{MoustosAbah2025B,VSRQHE}.  However, the analytical derivation of the generalized Carnot efficiency is still missing in the literature. Moreover, existing studies have been limited to adiabatic driving, leaving the role of nonadiabatic dynamics unexplored. In this work, we address this gap by deriving analytical expressions for the generalized efficiency bounds of relativistic quantum heat engines in both adiabatic and nonadiabatic regimes.
Specifically, we analyze the performance of a relativistic quantum Otto cycle with a time-dependent harmonic oscillator as the working medium. The nonadiabatic regime is treated in the sudden-switch limit, corresponding to instantaneous compression and expansion strokes. For both regimes, we obtain closed-form expressions for the upper bounds on the efficiency.

 The paper is organized as follows. In Sec. II, we present the model of the relativistic quantum Otto cycle with a harmonic oscillator as the working medium. In Sec. III, we derive an analytical expression for the generalized Carnot efficiency in the adiabatic regime. Section IV addresses the sudden-switch (nonadiabatic) regime and provides the corresponding generalized efficiency bound. Finally, we conclude in Sec. V.

\begin{figure}[ ]
 \begin{center}
\includegraphics[width=8.6cm]{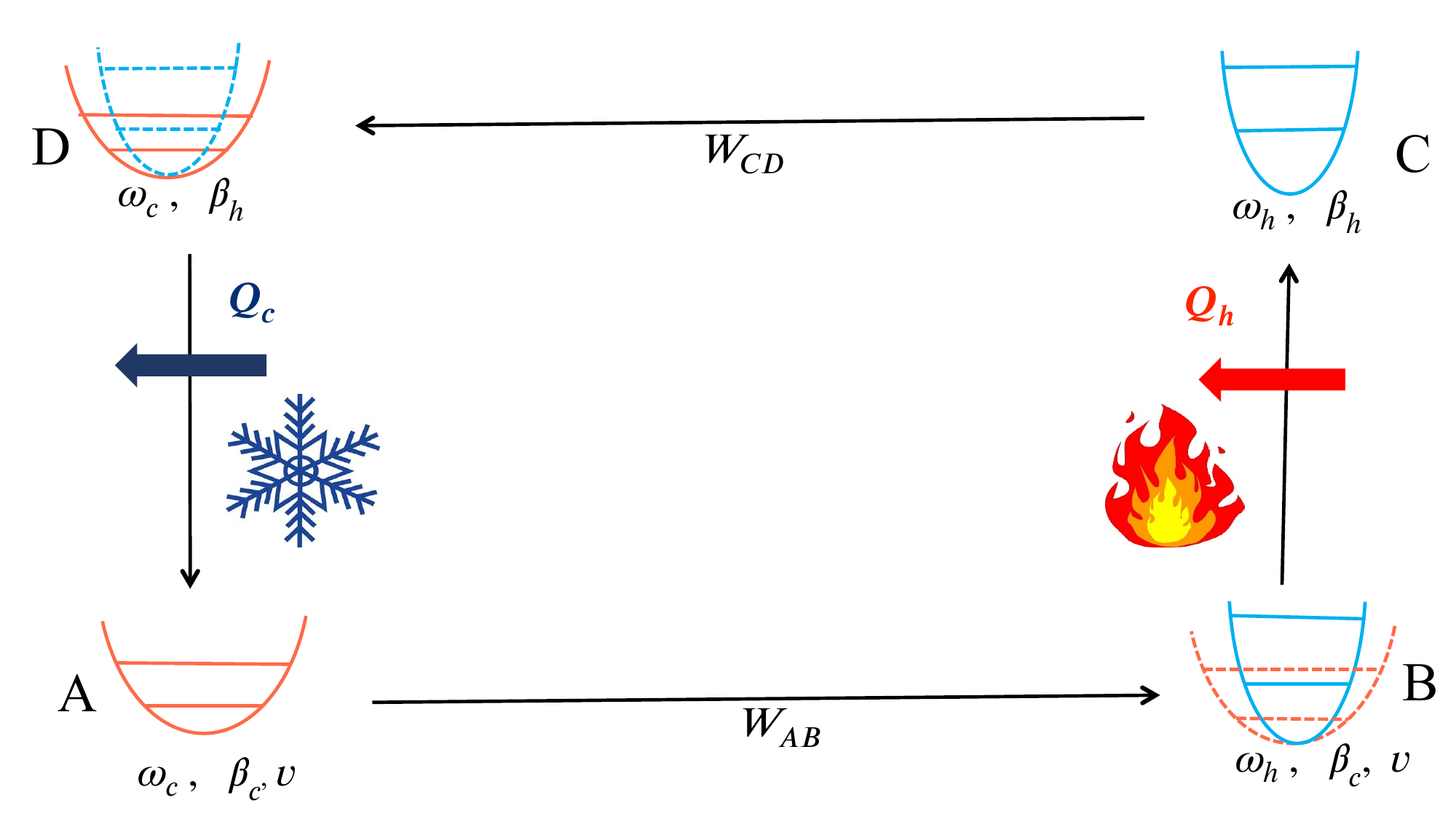}
 \end{center}
\caption{Schematics of quantum Otto cycle employing time-dependent harmonic oscillator as the working fluid. }
\end{figure}


%

\section{Relativistic Quantum Otto cycle }
We construct a relativistic quantum Otto cycle in which the working medium is a Unruh--DeWitt (UDW) \cite{Unruh1976,Dewitt1979} detector modeled as a harmonic oscillator \cite{Brown2013,Bruschi2013}. 
The cycle consists of four strokes: two adiabatic frequency-modulation strokes and two isochoric thermalization strokes. 
The detector interacts with scalar field reservoirs through the UDW coupling
\begin{equation}
    H_{\text{int}}(t) = \lambda \, m(t) \, \phi\!\left[x(\tau)\right],
\end{equation}
where $m(t) = a e^{-i\omega t} + a^\dagger e^{i\omega t}$ is the monopole operator of the oscillator, 
and $\phi[x(\tau)]$ is the field operator evaluated along the detector's worldline $x(\tau)$. 
In our setup, the oscillator remains stationary while coupled to the hot bath, 
$ x^\mu(\tau) = (\tau,0,0,0)$, whereas during interaction with the cold bath it moves uniformly with constant velocity $v$, following the worldline 
    $x^\mu(\tau) = \big(\gamma \tau, v \gamma \tau, 0,0\big)$, 
    where $\gamma = 1/\sqrt{1-v^2}$.

 The four steps of the quantum Otto cycle are explained as follows \cite{Lutz2012,LutzEPL}:
(1) Adiabatic compression $A \longrightarrow B$: To begin with, we consider the harmonic oscillator initially at thermal state with the cold bath while moving at constant velocity $v$. In this stage, the detector is isolated from the reservoirs, and the oscillator frequency is increased from   $\omega_c$ to $\omega_h$. Since the evolution is unitary, no heat is exchanged with the baths, and only work is performed on the system.
(2) Hot isochore $B\longrightarrow C$: During this stage, the oscillator is brought into contact with the hot reservoir at inverse temperature $\beta_h$  while its frequency is kept fixed at $\omega_h$. 
At the end of the hot isochoric stage, the system relaxes into a thermal state characterized by a mean photon number $\langle n_h\rangle=1/(e^{\beta_h\omega_h}-1)$ (we set $\hbar=k_B=1$). 
%
(3) Adiabatic expansion $C \longrightarrow D$: In this stage, the oscillator evolves in isolation while its frequency is decreased unitarily from  $\omega_h$   back to its initial value $\omega_c$. No heat exchange occurs, and work is performed by the system.
(4) Cold isochore $D\longrightarrow A$: In the final stroke, the oscillator—now moving with constant velocity $v$—is coupled to the cold reservoir at inverse temperature $\beta_c$ ($\beta_c>\beta_h$). During this interaction, the system exchanges heat with the cold bath and relaxes back to its initial thermal state $A$, thereby completing the cycle. Since the qubit is moving relativistically here, the system is thermalized to a state with mean photon number: $\langle n(\beta_c,v) = \ln[{(1-e^{-\beta_c\omega_c\gamma(1+v)})/((1-e^{-\beta_c\omega_c\gamma(1-v)})}]/2\gamma\,v\beta_c \omega_c$, which explicitly capture the impact of relativistic motion via the   velocity $v$ \cite{PapadatosA,PapadatosB,VSRQHE}.

The average energies, $\langle H\rangle=(\langle n(\beta_i,v)\rangle +1/2)\omega_i$ (where $i=c,h$ and $v=0$ for $i=h$), of the oscillator at the four stages of the cycle  can be written in the form \cite{Lutz2014}:

\begin{equation}
\langle H\rangle_A = 
\frac{\sqrt{1-v^{2}}}{2\beta_c v}\,
\ln\!\left[
\frac{\sinh\!\Big(\frac{\beta_c \omega_c}{2}\sqrt{\frac{1+v}{1-v}}\Big)}
{\sinh\!\Big(\frac{\beta_c \omega_c}{2}\sqrt{\frac{1-v}{1+v}}\Big)}
\right],
\end{equation}
\vspace{1mm}
\begin{equation}
\langle H\rangle_B =\frac{\sqrt{1-v^{2}}}{2\beta_c v}\,
\ln\!\left[
\frac{\sinh\!\Big(\frac{\beta_c \omega_c}{2}\sqrt{\frac{1+v}{1-v}}\Big)}
{\sinh\!\Big(\frac{\beta_c \omega_c}{2}\sqrt{\frac{1-v}{1+v}}\Big)}
\right]\frac{\omega_h}{\omega_c}\lambda ,
\end{equation}
\begin{equation}
\langle H_C\rangle = \frac{\omega_h}{2}\coth\left(\frac{\beta_h\omega_h}{2}\right),
\end{equation}
\begin{equation}
\langle H_D\rangle = \frac{\omega_c}{2}\lambda \coth\left(\frac{\beta_h\omega_h}{2}\right) ,
\end{equation}
where $\lambda$ is the dimensionless adiabaticity parameter of the dynamics, determined by the specific type of frequency modulation (more details can be found in \cite{Husimi,Deffner2008}). In general, $\lambda \geq 1$, and its explicit form is given by
\begin{equation}
\lambda = \frac{1}{2\omega_c \omega_h} 
\Big\{ \omega_c^2 \big[ \omega_h^2 X(t)^2 + \dot{X}(t)^2 \big] 
+ \big[ \omega_h^2 Y(t)^2 + \dot{Y}(t)^2 \big] \Big\},
\end{equation}
where $X(t)$ and $Y(t)$ are solutions of the equation $d^2X/dt^2 + \omega^2(t)X = 0$, with initial conditions $X(0)=0$, $\dot{X}(0)=1$, $Y(0)=1$, and $\dot{Y}(0)=0$  \cite{Husimi,Deffner2008}.

The mean work and heat exchanges during the work strokes and the hot isochoric stroke can be evaluated as%
 \begin{widetext}
 \begin{equation}
    W_{AB} =\langle H \rangle_B-\langle H \rangle_A  
 = \frac{\sqrt{1-v^{2}}}{2\beta_c \omega_c v}\,
\ln\!\left[
\frac{\sinh\!\Big(\frac{\beta_c \omega_c}{2}\sqrt{\frac{1+v}{1-v}}\Big)}
{\sinh\!\Big(\frac{\beta_c \omega_c}{2}\sqrt{\frac{1-v}{1+v}}\Big)}
\right](\omega_h \lambda-\omega_c) , 
  \\ 
  \label{wab}
\end{equation}
\begin{equation}
 W_{CD}= \langle H \rangle_D-\langle H \rangle_C  
= \frac{1}{2}\coth\left(\frac{\beta_h\omega_h}{2}\right)(\omega_c \lambda - \omega_h),
  \label{wcd}   
  \end{equation}
  \begin{equation}
 Q_h = \langle H \rangle_C-\langle H \rangle_B 
 = \frac{\omega_h}{2} \left( \coth\left(\frac{\beta_h\omega_h}{2}\right) - \frac{\sqrt{1-v^{2}}}{2\beta_c \omega_c v}\,
\lambda \,\ln\!\left[
\frac{\sinh\!\Big(\frac{\beta_c \omega_c}{2}\sqrt{\frac{1+v}{1-v}}\Big)}
{\sinh\!\Big(\frac{\beta_c \omega_c}{2}\sqrt{\frac{1-v}{1+v}}\Big)}
\right]  \right). \label{qh}
 \end{equation}
Here, we are employing a sign convention in which heat is absorbed (rejected) from (to) the
reservoir is positive (negative).
After one complete cycle, the working fluid returns to its initial state, so the net work performed on the system is determined by the first law of thermodynamics as $W_{\rm total}=-(  Q_h + Q_c)=-(W_{AB}+W_{CD})$. Work is extracted from the engine when   $W_{\rm ext}=-W_{\rm total}=W_{AB}+W_{CD}>0$. Using Eqs.~(\ref{wab})–(\ref{qh}), the final expressions for the extracted work, $W_{\rm ext}$, and the efficiency, $\eta =  W_{\rm ext}/Q_h$, are obtained as
%
%
\begin{equation}
W_{\rm ext} = \frac{\omega_h - \lambda \,\omega_c}{2 v \,\beta_c \,\omega_c}
\left[
  v \,\beta_c \,\omega_c \, \coth\!\left(\tfrac{\beta_h \omega_h}{2}\right)
  - \sqrt{1-v^2}\,
    \ln\!\left(
      \csch\!\left(\sqrt{ \frac{1-v}{1+v}}\,\frac{\beta_c \omega_c}{2}\right)
      \sinh\!\left(\frac{1}{2}\sqrt{\frac{1+v}{1-v}}\,\beta_c \omega_c\right)
    \right),
\right] \label{workgen}
\end{equation}

\begin{eqnarray}
\eta =  \frac{W_{\rm ext}}{Q_h}=
\frac{
  v\,\beta_c \omega_c\,(-\lambda \omega_c + \omega_h)\,\coth\!\left(\frac{\beta_h \omega_h}{2}\right)
  + \sqrt{1-v^2}\,(\omega_c - \lambda \omega_h)\,
    \ln\!\Big[
      \operatorname{csch}\!\big(\tfrac{1}{2}\sqrt{\tfrac{1-v}{1+v}}\,\beta_c \omega_c\big)\,
      \sinh\!\big(\tfrac{1}{2}\sqrt{\tfrac{1+v}{1-v}}\,\beta_c \omega_c\big)
    \Big]
}{
  \omega_h\!\left(
    v\,\beta_c \omega_c\,\coth\!\left(\frac{\beta_h \omega_h}{2}\right)
    - \sqrt{1-v^2}\,\lambda\,
      \ln\!\Big[
        \operatorname{csch}\!\big(\tfrac{1}{2}\sqrt{\tfrac{1-v}{1+v}}\,\beta_c \omega_c\big)\,
        \sinh\!\big(\tfrac{1}{2}\sqrt{\tfrac{1+v}{1-v}}\,\beta_c \omega_c\big)
      \Big]
  \right)
}. \label{efficiency}
\end{eqnarray}
 \end{widetext}
\section{Generalized Carnot bound on the efficicency for adiabatic driving}
We are primarily interested in obtaining the analytical expression for generalized upper bound on the efficiency, which can be obtained in two limiting cases of the driving protocol: (i) adiabatic driving, where $\lambda=1$, and (ii) the sudden-switch protocol, for which $\lambda=(\omega_c^2+\omega_h^2)/2\omega_c\omega_h$. 
Starting with the first limiting case of adiabatic driving, the  efficiency in Eq. (\ref{efficiency}) takes the familiar form
\begin{equation}
    \eta  = 1-\frac{\omega_c}{\omega_h}. \label{efficiency2}
\end{equation}
Relativistic effects do not alter the functional form of the efficiency, but they do influence the amount of work extracted, as will be shown in the following analysis.  Henceforth, for further analysis, we will assume the high-temperature regime for which $\beta_{h (c)}\omega_{h(c)}\ll 1$. In this limit, the expression for extracted work, $  W_{\rm ext}  = -(W_{\rm AB}+W_{\rm CD})$, reads as
\begin{equation}
   W_{\rm ext}
   = \frac{(\omega_h - \omega_c)\,\big(\beta_c \omega_c -  \beta_h \omega_h f(v))}
{\beta_c \beta_h \omega_c \omega_h},\label{Wext}
\end{equation}
where we have introduce $f(v)= \sqrt{1-v^2}\ln{[(1+v)/(1-v)]}/2v$.
 Equation~(\ref{Wext}) can be recast into a compact form by introducing the frequency ratio $z=\omega_c/\omega_h$ and the temperature ratio $\tau=\beta_h/\beta_c$:
\begin{equation}
 W_{\rm ext} = \frac{1-z}{\beta_c\tau\,z}\left(   z-\tau   f(v)  \right). \label{Wz}
\end{equation}
 %
 \begin{equation}
 W_{\rm ext} = \frac{1-z}{\beta_c\tau\,z}\left[   z-\tau  \frac{\sqrt{1-v^{2}}}{2v} \, \ln\!\left(\frac{1+v}{1-v}\right)  \right]. \label{Wz}
\end{equation}
The positive--work condition $W_{\rm ext}\!\ge 0$ implies
\begin{equation}
    z \;\ge\; \tau\, f(v), \qquad 
    f(v)\equiv \frac{\sqrt{1-v^{2}}}{2v}\,\ln\!\frac{1+v}{1-v}. \label{PWC}
\end{equation}
Combining Eq.(\ref{efficiency2}) with the PWC condition in Eq.(\ref{PWC}), we obtain the following analytical expression for the upper bound on the efficiency of the relativistic engine:
    \begin{equation}
          \eta  \leq 1   - \frac{\beta_h}{\beta_c} \frac{2v}{1-v^{2}} \, \ln\!\left(\frac{1-v}{1+v}\right) \equiv \eta^{\rm gen}_{\rm C}(v). \label{upperbound}
    \end{equation}
Note that $\eta_C \leq \eta^{\rm gen}_C(v)$, with equality at $v=0$, while in the ultra-relativistic limit $v \to 1$ the generalized bound approaches unity, $\eta^{\rm gen}_C(v)\to 1$. This generalized bound was first proposed in Ref.~\cite{VSRQHE}, where it was deduced numerically from efficiency–power tradeoff characteristics in steady-state heat engines. In contrast, here we provide, for the first time, a closed-form analytical derivation of this bound. Since the same efficiency emerges both in steady-state and cyclic relativistic engines, we identify it as the generalized Carnot bound governing the efficiency of all relativistic heat engines.

The form of $\eta^{\rm gen}_{\rm C}(v)$ can be further understood by assigning an effective temperature to the moving cold reservoir through the concept of a  directional temperature\cite{Costa1995,Landsberg1996} perceived in the qubit’s rest frame,
\begin{equation}
T^{\theta}_{\rm eff} = \frac{T\sqrt{1-v^2}}{1-v\cos{\theta}}, \label{Tdirection}
\end{equation}
where $\theta$ denotes the angle between the motion axis and the line of sight. Averaging over the full solid angle $4\pi$   gives $\big\langle (1 - v\cos\theta)^{-1} \big\rangle
=  \ln{[(1+v)/(1-v)]/2v}$, 
which leads to an isotropic effective temperature for the moving cold reservoir,   $T_c^{\rm eff} = T_c \sqrt{1-v^2}\ln{ [(1+v)/(1-v)]}/2v.$, and, in turn, yields the familiar Carnot form $\eta^{\rm gen}_{C}(v)=1-T^{\rm eff}_{c}/T_h$.

\begin{figure}
    \centering
    \includegraphics[width=1\linewidth]{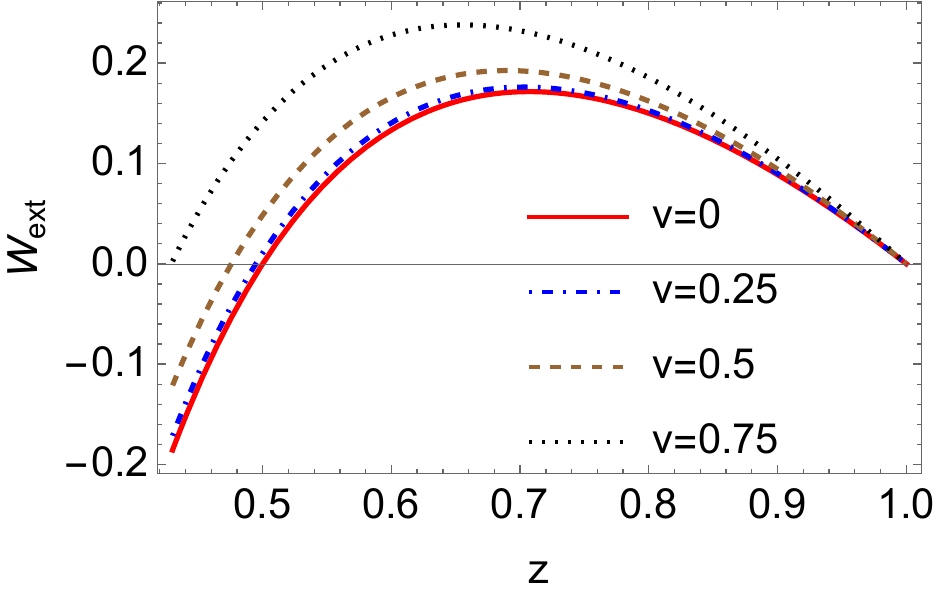}
    \caption{Work output (Eq. (\ref{Wz})) as a function of compression ratio $z$ for different fixed values of $v$. Here, $\tau=1/2$. }
    \label{workvsz}
\end{figure}

\begin{figure}
    \centering
    \includegraphics[width=1\linewidth]{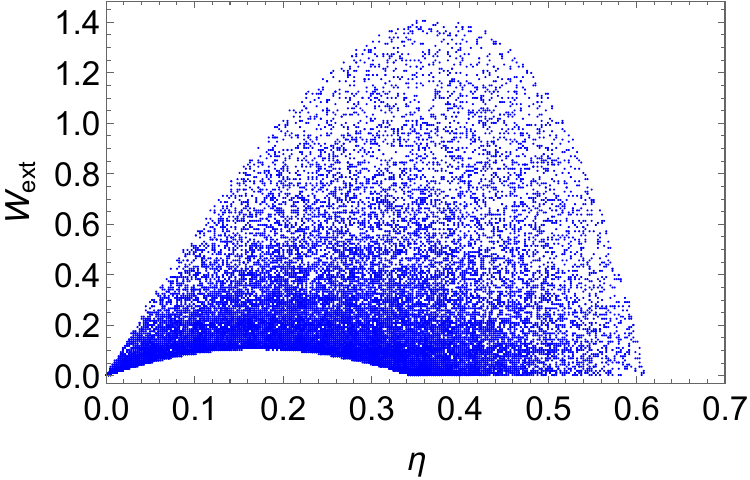}
    \caption{Scatter plot of extracted work $W_{\rm ext}$ [Eq.(\ref{workgen})] versus efficiency $\eta$ [Eq.(\ref{efficiency2})]. The outer envelope exhibits the characteristic efficiency–power trade-off, and all points lie strictly below the maximum efficiency predicted by Eq.~(\ref{upperbound}), providing clear evidence that it constitutes a universal upper bound. Parameters are fixed as $\beta_h = 1/10$ and $\beta_c = 1/5$; for $v=0.85$ the generalized Carnot bound is $\eta^{\rm gen}_C = 0.612$. Frequencies are randomly sampled from $\omega_c \in (0,30)$ and $\omega_h \in (0,60)$ } 
    \label{scatterplot}
\end{figure}

Although the generalized Carnot bound in Eq.(\ref{upperbound}) was derived in the high-temperature regime, we argue that it in fact constitutes a universal upper bound, valid across all operational regimes. To support this claim, Fig.\ref{scatterplot} shows a scatter plot of the extracted work $W_{\rm ext}$ [Eq.(\ref{workgen})] versus efficiency $\eta$ [Eq.(\ref{efficiency2})]. The outer envelope of the data points exhibits the characteristic efficiency–power trade-off, and, importantly, all points lie strictly below the maximum efficiency defined in Eq.(\ref{upperbound}), providing clear evidence that Eq.(\ref{upperbound}) serves as a genuine and universal upper bound.

Having discussed the effect of relativistic motion on the efficiency bound, we will shed more light on the effects of relativistic motion on the positive work extraction condition.  This differs from the stationary--bath case, where $z\ge \tau$. 
The function $f(v)$ is even in $v$, satisfies $0< f(v)\le 1$ with $f(0)=1$, 
and decreases monotonically on $0<v<1$ (indeed $f(v)\to 0$ as $v\to1^{-}$). 
Hence, motion relaxes the work--extraction threshold: $z$ can be smaller than $\tau$ while still yielding $W_{\rm ext}\ge0$.  
Figure~\ref{workvsz} illustrates these trends, where we plot Eq.~(\ref{Wz}) versus $z$ for a fixed temperature ratio $\tau=1/2$.  Each curve $W_{\rm ext}(z)$ at fixed $v$ crosses zero at
\begin{equation}
    z_{\min}(v)=\tau\,f(v), \label{eq:work-threshold}
\end{equation}
and the zero--crossing shifts monotonically leftward with increasing $v$, in line with Eq.~\eqref{eq:work-threshold}. For a fixed $z$ in the plotted range, the value of $W_{\rm ext}$ increases systematically with $v$. This can be understood as follows.  In the qubit’s rest frame, the bath spectrum no longer looks thermal. Doppler reshaping effectively lowers the cold bath temperature, which enlarges the temperature difference between the two reservoirs. This enhanced gradient broadens the parameter window in $(\omega_c,\omega_h)$ space where the engine can operate, thereby allowing greater work extraction.

\begin{figure}
    \centering
    \includegraphics[width=1\linewidth]{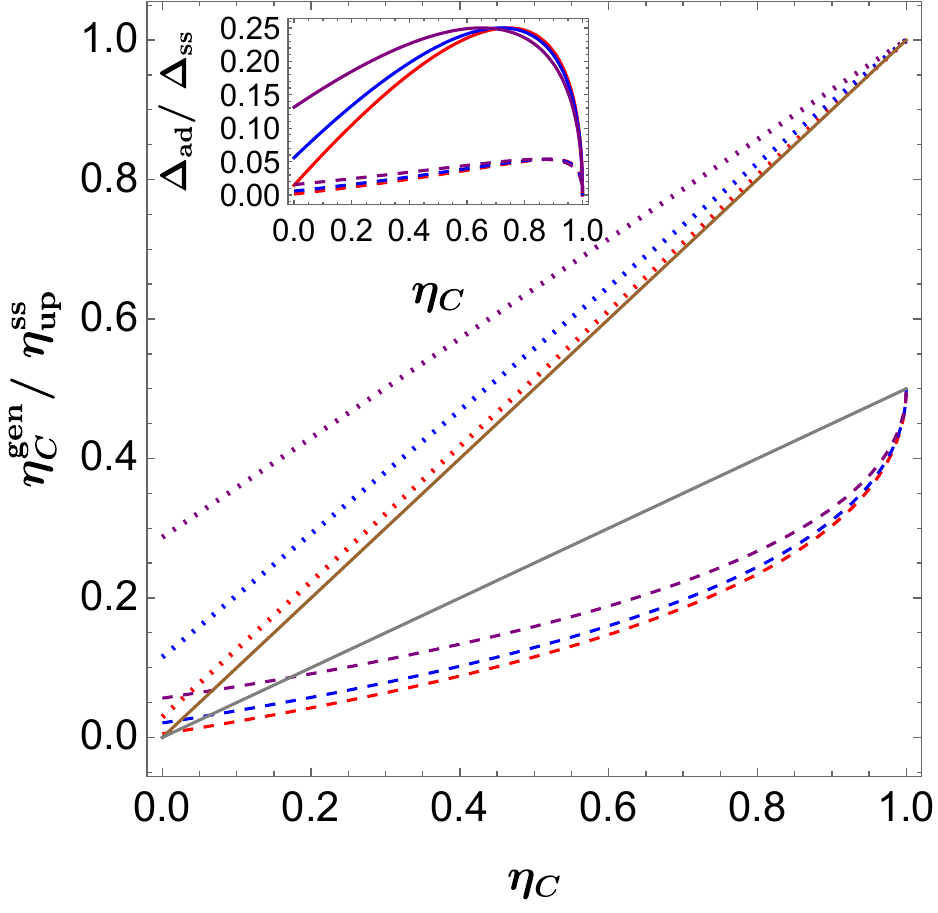}
    \caption{Efficiency bounds for the relativistic quantum Otto engine. The main panel shows the generalized Carnot efficiency, $\eta^{\rm gen}_{C}(v)$ [Eq.(\ref{upperbound})], plotted as a function of the standard Carnot efficiency $\eta_C$ for several fixed velocities $v$ (dotted curves). All dotted curves lie above the conventional Carnot bound (solid brown line). The corresponding sudden-switch bounds [Eq.(\ref{etaupss})] are shown as dashed curves of the same color. From bottom to top, the velocities are $v=0.4, 0.7, 0.9$.
    Inset: the difference $\Delta_{\rm ad}=\eta^{\rm gen}_{C}(v)-\eta^{\rm MW}(v)$ between the generalized Carnot efficiency and the efficiency at maximum work. The solid curves show that $\Delta{\rm ad}$ remains strictly positive, confirming that  $\eta^{\rm MW}(v)$  is always bounded from above by $\eta^{\rm gen}_{C}(v)$. The lower set of curve (dashed curves)  in the same color correspond to the sudden-switch case: $\Delta_{\rm ss}=10(\eta^{\rm SS}_{\rm up}-\eta^{\rm SS}_{\rm MW}$).    } 
    \label{alleffs}
\end{figure}

Now, for completeness's sake, now we discuss efficiency at maximum work. Optimization of Eq. (\ref{Wz})   with respect to $z$ the following form of efficiency at maximum work,
\begin{equation}
    \eta^{\rm MW}(v) = 1 -  \sqrt{1-\tau \,f(v)}, 
\end{equation}
which reduces to the well-known Curzon-Ahlborn form in the limit $v\rightarrow0$. While $\eta^{\rm MW}(v)$ can exceed the standard Carnot efficiency for certain parameter ranges, it always remains bounded from above by the generalized Carnot limit given in Eq.(\ref{upperbound}). This is illustrated in the inset of Fig.\ref{alleffs}, where we plot the difference  $\Delta{\rm ad}=\eta^{\rm gen}_{C}(v)-\eta^{\rm MW}(v)$, which is always positive (solid curves). In the main figure,  $\eta^{\rm gen}{C}(v)$ from Eq.~(\ref{upperbound}) from Eq.~(\ref{upperbound}) is plotted as a function of $\eta_C$for several fixed values of $v$ (dotted curves), all of which lie above the standard Carnot bound (solid brown curve).

\section{Performance bounds for sudden switch protocol}
Next, we will discuss the performance bounds of the relativistic heat engine for the sudden-switch driving protocol, which is another analytically solvable case. In the sudden-switch regime, $\lambda=(\omega_c^2+\omega_h^2)/2\omega_c\omega_h$. For $\lambda=\omega_c^2+\omega_h^2/2\omega_c\omega_h$, Eqs. (\ref{workgen}) and (\ref{efficiency}) take the following forms, respectively:
\begin{equation}
W^{\rm SS}_{\rm ext}=\frac{\omega_h^{2}-\omega_c^{2}}{4 v\,\beta_c\,\omega_c^{2}\,\omega_h}
\left[
  v\,\beta_c\,\omega_c^{2}\,\coth\!\left(\frac{\beta_h \omega_h}{2}\right)
  - \omega_h A
\right], \label{workss}
\end{equation}
\begin{equation}
    \eta^{\rm SS} = \frac{(\omega_h^{2}-\omega_c^{2})
\left(
  v\,\beta_c\,\omega_c^{2}\,\coth\!\left(\frac{\beta_h \omega_h}{2}\right)
  - \omega_h A
\right)}
{\omega_h\left(
  2 v\,\beta_c\,\omega_c^{2}\,\omega_h\,\coth\!\left(\frac{\beta_h \omega_h}{2}\right)
  - (\omega_c^{2}+\omega_h^{2})\,A
\right)}, \label{etass}
\end{equation}
where we have introduced the notation 
$A=\sqrt{1-v^2}\ln\!\left[
  \csch\!\left(\tfrac{1}{2}\sqrt{\tfrac{1-v}{1+v}}\,\beta_c \omega_c\right)
  \,\sinh\!\left(\tfrac{1}{2}\sqrt{\tfrac{1+v}{1-v}}\,\beta_c \omega_c\right)
\right]$.
After some algebra, Eq. (\ref{etass}) can be recasted into the form 
\begin{eqnarray}
     \eta^{\rm SS} &=&
     \left[
                \frac{2}{1-\frac{\omega_c^2}{\omega_h^2} }+ \frac{A}{v\,\beta_c\coth{\left( \frac{\beta_h\omega_h}{2}\right)}\omega_c^2\omega_h-A\omega_h^2}
     \right]^{-1} \nonumber
     \\
     &=&  \left[
                    \frac{2}{1-\frac{\omega_c^2}{\omega_h^2} } + \frac{1}{\omega_c^2\omega_h}
                  \Bigg\{    \frac{v\,\beta_c \coth{\left( \frac{\beta_h\omega_h}{2}\right)}}{A}   - \frac{\omega_h}{\omega_c^2}
                       \Bigg\}^{-1}
     \right]^{-1} \nonumber
     \\
     & \equiv & \left(\frac{2}{\Delta_1} +   \frac{1}{\omega_c^2\omega_h} \Delta_2^{-1}       \right)^{-1}. \label{etass2}
\end{eqnarray}
Now, imposing the positive–work condition (PWC) from Eq.~(\ref{workss}), 
$W^{\rm SS}_{\rm ext}>0$, we obtain
\begin{align}
v\,\beta_c\,\omega_c^{2}\,\coth\!\left(\frac{\beta_h \omega_h}{2}\right)-\omega_h A \;>\; 0, 
\end{align}
which implies that $\Delta_2> 0$, which further results in $1/\Delta_2> 0$. Further, as $0<\omega_c<\omega_h$, $\Delta_1=1-\omega_c^2/\omega_h^2 \in (0,1)$. From Eq. (\ref{etass2}), it follows that
\begin{equation}
    \eta^{ss} < \frac{\Delta_1}{2} \qquad \text{and}\qquad \eta^{\rm SS} < \Delta_2. \label{DD}
\end{equation}
Since $0<\Delta_1<1$ in Eq. (\ref{DD}), this immediately yields our second main result:
\begin{equation}
    \eta^{\rm SS} <\frac{1}{2}. \label{etahalf}
\end{equation}
 The result is striking: even in the ultra-relativistic limit $v\rightarrow1$, the engine efficiency cannot exceed  1/2. This sharply contrasts with quasi-static analysis in Sec.~II, where the efficiency can asymptotically approach unity in the limit $v\rightarrow1$. The difference originates in the strongly nonadiabatic character of the sudden-switch protocol: an instantaneous quench of the oscillator frequency generates transitions and coherences in the instantaneous energy basis, storing parasitic internal energy in the working medium. During the subsequent isochores, this excess is dissipated to the reservoirs—an effect known as inner (quantum) friction \cite{Rezek2010,Plastina2014,Rezek2017,Feldmann2000,Feldmann2002,OzgurFriction}. While relativistic Doppler reshaping enhances the thermodynamic bias, it cannot overcome these frictional losses, which ultimately enforce the 
1/2 efficiency cap in the sudden-switch regime.

\subsection{Upper bound on the efficiency}
 \begin{figure}
      \centering
      \includegraphics[width=1\linewidth]{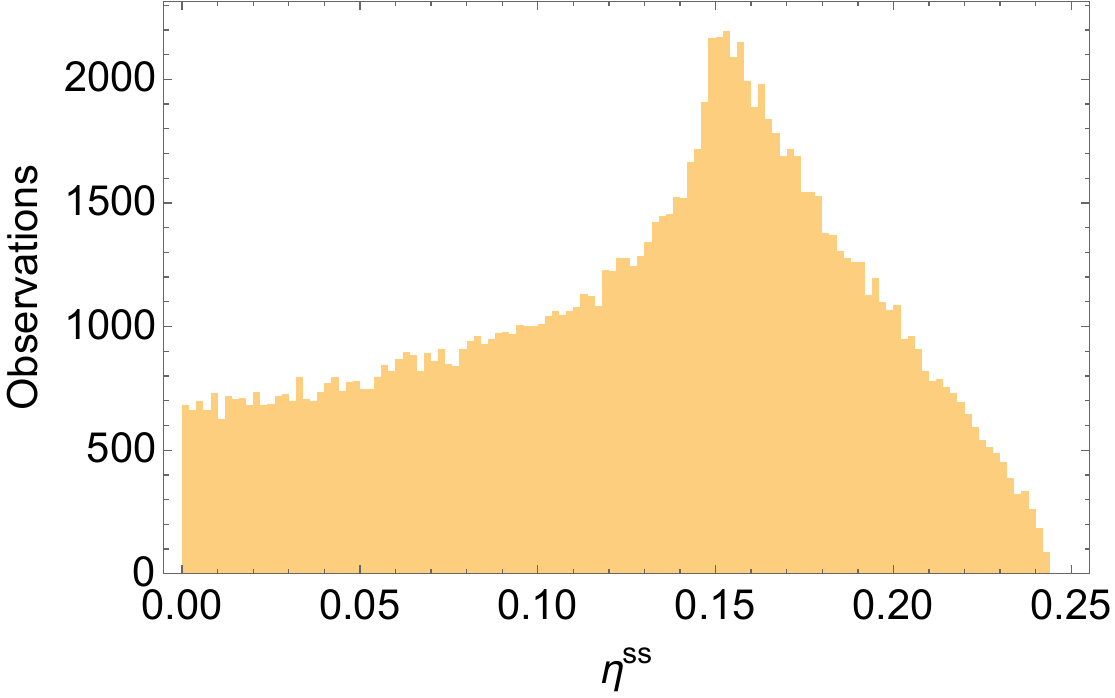}
      \caption
      {
      Histogram of the sampled values of $\eta^{\rm SS}$ given in Eq. (\ref{etass}) obtained from random sampling over the parameter space  $(\omega_c, \omega_h)$. Frequencies are drawn from uniform distributions $\omega_c\in[0, 20]$ and $\omega_h\in [0, 40]$, with fixed parameters $\beta_h=1/20$, $\beta_c=1/5$, and $v=0.9$. A total of $10^6$ 
  random events are generated. The distribution clearly shows that all sampled efficiencies remain below the analytic upper bound $\eta^{\rm SS}_{\rm up}=0.24366$ given by Eq. (\ref{etaupss}), thereby establishing   $\eta^{\rm SS}_{\rm up}$ as the true upper limit of efficiency across all operational regimes.
 }
      \label{histogram}
  \end{figure}
To obtain closed-form analytic expressions for the efficiency, we confine ourselves to the high-temperature regime \cite{Kosloff1984,UzdinEPL,VJ2019}, where $\coth(\beta_i\omega_i/2)\approx 2/(\beta_i\omega_i)$ ($i=c, h$). Under this approximation, the expressions for extracted work $W^{\rm SS}_{\rm ext}$ [Eq.(\ref{workss})] and the efficiency $\eta^{\rm SS}$ [Eq.(\ref{etass})] reduce to:
\begin{eqnarray}
   W^{\rm SS}_{\rm ext} =   &=& \frac{(1-z^2) \left[z^2 -\tau\, f(v) \right]}{2z^2\beta_h}, \label{workssHT}
\\
\eta^{\rm SS} &=& \frac{(z^2-1)[z^2  - \tau\,f(v)]}{\tau - z^2[2-\tau\,f(v)]}, \label{etassHT}  
\end{eqnarray}
where $z=\omega_c/\omega_h$, $\tau=\beta_h/\beta_c=1-\eta_C$ and $f(v)=\sqrt{1-v^2}\ln[(1+v)/(1-v)]/2v$.  In order to obtain the analytic expression of the efficiency upper bound, we again use the PWC in Eq. (\ref{workssHT}),  $  W^{\rm SS}_{\rm ext}>0$, which implies that 
\begin{equation}
    z^2 > \tau \frac{\sqrt{1-v^2}}{2v}\ln{\left[ \frac{1+v}{1-v}   \right]}. \label{PWCss}
\end{equation}
From the efficiency expression in Eq. (\ref{etassHT}), $z^2$  can be expressed in terms of $\eta$ and $\tau$ as:
\begin{widetext}
    \begin{equation}
   z^{2}
= \frac{1}{2}\!\left(
1 - 2\eta +f(v)\tau + \eta\,\tau \,f(v)
- \sqrt{\,4f(v)(\eta-1)\tau + \bigl(1 +f(v)\tau + \eta(-2 +f(v)\tau)\bigr)^{2}}
\right).
\end{equation}
\end{widetext}
Substituting the above expression for $z^2$ into Eq. (\ref{PWCss}) and solving the resulting inequality yields the following upper bound on the engine’s efficiency in terms of $\eta_C:$
\begin{equation}
\eta^{\rm SS}_{\rm up} = \frac{1 -f(v)\,(1-\eta_C)}{\bigl(\sqrt{2}+\sqrt{f(v)\,(1-\eta_C)}\bigr)^{2}} <\frac{1}{2}. \label{etaupss}
 \end{equation}
This is our third main result. Notably, the derived bound is universal in the sense that it does not depend on the specific model parameters but only on the reservoir characteristics, namely $v$ and $\eta_C$	
  (or equivalently $\tau$). In the ultra-relativistic regime limit $v\rightarrow 1$, the upper efficiency approaches $\eta^{\rm SS}_{\rm up}\rightarrow 1/2$.  This recovers our earlier result [Eq. (\ref{etahalf})] that the maximum efficiency achievable by our engine is 1/2 only. We have plotted Eq.(\ref{etaupss}) as a function of $\eta_C$ in Fig.\ref{alleffs} for different fixed values of velocity $v$ (lower set of curves, shown as dashed lines). These curves converge to $1/2$ as $\eta_C \to 1$, and they lie well below their corresponding adiabatic counterparts (solid curves of the same color). This contrast clearly illustrates the detrimental impact of quantum friction on the engine’s performance bounds.Nevertheless, relativistic motion of the working medium can still be harnessed to surpass the standard Carnot bound: as seen in Fig.~\ref{alleffs}, the dashed blue and purple curves exceed the Carnot efficiency (solid brown curve) for small $\eta_C$. However, with increasing $\eta_C$, the advantage diminishes and the curves eventually drop below $\eta_C$, even falling beneath $\eta_C/2$ (solid gray curve) for higher values of $\eta_C$.

  To confirm that the upper bound obtained in Eq. (\ref{etaupss}) is not restricted to the high-temperature regime but in fact holds across all temperature regimes, we plot in Fig. (\ref{histogram}) a histogram of the general efficiency [Eq. (\ref{etass})] obtained from random sampling over the parameter space $(\omega_c, \omega_h)$. For the chosen set of parameters, every sampled efficiency value lies strictly below the bound of Eq. (\ref{etaupss}), thereby confirming that this expression provides the genuine upper limit on the efficiency of the relativistic engine in the sudden-switch regime.

Next, we examine the efficiency at maximum work. Optimizing Eq. (\ref{workssHT}) with respect to the compression ratio 
z
z gives the optimal value $z^*=\tau^{1/4}$. Substituting this result into Eq. (\ref{etassHT}) yields the following expression for the efficiency at maximum work:
\begin{equation}
    \eta^{\rm SS}_{\rm MW} = \frac{2 +f(v)\,\tau - 3\sqrt{f(v)\,\tau}}{4 -f(v)\,\tau}, \label{etassMW}
\end{equation}
which reduces to the Rezek–Kosloff efficiency at maximum work, $\eta_{\rm RK}=(1-\sqrt{\tau})/(2+\sqrt{\tau})$, originally derived in Ref.~\cite{Rezek2006}. To highlight the difference between $\eta^{\rm SS}_{\rm up}$ and $\eta^{\rm SS}_{\rm MW}$, we plot their deviation, $\Delta_{\rm SS}=\eta^{\rm SS}{\rm up}-\eta^{\rm SS}{\rm MW}$, as dashed curves in the inset of Fig.\ref{alleffs}. For visual clarity, $\Delta_{\rm SS}$ is multiplied by a factor of 10; nevertheless, the difference remains much smaller than the corresponding adiabatic counterpart $\Delta_{\rm ad}$. This indicates that in the sudden-switch regime, the engine’s performance is largely governed by frictional effects, rendering the distinction between the maximum work and maximum efficiency points negligible.

\section{Conclusions} 
We have investigated the thermodynamic performance of a relativistic quantum Otto engine with a harmonic oscillator as the working medium, demonstrating how relativistic motion   can enhance the engine’s efficiency. In particular, we derived an analytic expression for the generalized Carnot bound in the adiabatic driving regime, showing that it can surpass the standard Carnot efficiency and even approach unit efficiency in the ultra-relativistic limit.
We then analyzed the performance of the engine under the sudden-switch driving protocol, where frictional effects arise from nonadiabatic excitations induced by instantaneous changes in the system’s parameters. In this regime, we explored the interplay between relativistic and frictional effects and found that, in sharp contrast to the adiabatic case, the efficiency is strongly suppressed: even in the ultra-relativistic limit, it cannot exceed $1/2$. Furthermore, we derived an analytical expression for the corresponding upper bound on efficiency, which serves as the sudden-switch counterpart of the generalized Carnot bound obtained for adiabatic driving. 

More importantly, this work provides the first systematic account of frictional effects in relativistic quantum heat engines, establishing analytical efficiency bounds beyond the adiabatic regime. Together, the analytic bounds derived here offer a benchmark for assessing the role of relativistic and frictional effects in future models of relativistic quantum heat engines.

 \section{Acknowledgements} 
This research was supported by individual KIAS Grants   No. PG096801 (V.S.) at the Korea Institute for Advanced Study.
  

%

\end{document}